# Dynamical Casimir effect for different geometries

**D A R Dalvit[1], F D Mazzitelli[2], and X Orsi Millán[2]**

[1] Theoretical Division, MS B213, Los Alamos National Laboratory, Los Alamos, NM 87545, USA

[2] Departamento de Fisica J.J. Giambiagi, Facultad de Ciencias Exactas y Naturales, Universidad de Buenos Aires, Ciudad Universitaria, Pabellón 1, 1428 Buenos Aires, Argentina

E-mail: `dalvit@lanl.gov, fmazzi@df.uba.ar, xorsi@df.uba.ar`

**Abstract.**
We consider the problem of motion-induced photon creation from quantum vacuum inside closed, perfectly conducting cavities with time-dependent geometries. These include one dimensional Fabry-Perrot resonators with Dirichlet or Neumann boundary conditions, three dimensional cylindrical waveguides, and a spherical shell. The number of Casimir TE, TM and TEM photons is computed. We also present a classical mechanical analogue of the one dimensional dynamical Casimir effect.

PACS numbers: 42.50.Lc, 42.40.Dv

*Keywords*: dynamical Casimir effect, motion induced radiation, cavity QED

## 1. Introduction

Conversion of zero-point quantum fluctuations into real particles occurs in quantum field theory under the influence of time-dependent external conditions. This effect is generally known as the dynamical Casimir effect (DCE). Examples range from cosmology, such as particle creation in curved spacetimes [1], to cavity QED, such as photons production in Fabry-Perrot cavities with moving mirrors [2, 3, 4, 5, 6, 7, 8, 9]. In this latter context, on which we will concentrate in this present contribution, the periodic modulation of the boundary conditions of the electromagnetic field resonantly excite particular modes of the field, transforming the initial vacuum state into a squeezed state, thereby creating real photons. Up to now no experimental verification of the effect has been reported in the literature. The main difficulties are the stringent requirements on the modulation frequency, amplitude, and optical quality factor of the cavity. There are a few ongoing experimental projects to demonstrate different variants of the dynamical Casimir effect. One of them is being carried out in the MIR experiment [10], and it is based on the idea that when a thin semiconductor layer inside a high Q cavity is periodically illuminated by laser pulses, its conductivity properties are periodically modulated, and this simulates an effecting moving mirror. Another project, under development at Dartmouth [11], aims at detecting atomic transitions of ultracold atomic samples induced by Casimir photons produced inside cavities with high-frequency nanoresonators.

In this paper we report on analytical methods to study the dynamical Casimir effect in different dimensions and different geometries, both for scalar fields and for the full electromagnetic field. We start in Section II by briefly reviewing the computation of motion-induced scalar particles in 1+1 dimensions. We show that Moore's formalism [2], originally used to treat Dirichlet boundary conditions, can be straightforwardly extended to Neumann boundary conditions. In order to gain some intuitive picture of the dynamical Casimir effect, we present in Section III a classical mechanical *simulator* of the 1+1 DCE. The full electromagnetic dynamical Casimir effect inside perfectly conducting cavities with time-dependent boundaries is reviewed in Section IV, where we deal with TEM modes in non-simply connected cavities, and in Section V, where TE and TM modes are considered in waveguides and in a spherical shell.

## 2. 1+1 dynamical Casimir effect: Dirichlet and Neumann boundary conditions

Let us consider a massless real scalar field in a one dimensional cavity with one end fixed at $z = 0$ and the other performing an oscillatory motion $L(t) = L_0[1 + \epsilon \sin(\Omega t)]$, where $\epsilon \ll 1$, $\Omega = q\pi/L_0$, and $q \in \mathbb{N}$. We shall assume that the oscillations begin at $t = 0$ and end at $t = T$. The scalar field $\phi(z,t)$ satisfies the wave equation $\Box \phi = 0$.

We first review the case of Dirichlet boundary conditions: $\phi(z = 0, t) = \phi(L(t), t) =$

0. In this case one can express the field inside the cavity as a sum of modes

$$\phi(z,t) = \sum_{k=1}^{\infty} \left[ a_k \psi_k(z,t) + a_k^\dagger \psi_k^*(z,t) \right], \quad (1)$$

where the mode functions $\psi_k(z,t)$ are positive frequency modes for $t < 0$, normalized according to the Klein-Gordon inner product, and $a_k$ and $a_k^\dagger$ are time-independent bosonic annihilation and creation operators, respectively. The field equation is automatically verified writing the modes in terms of Moore's function $R(t)$ [2] as

$$\psi_k(z,t) = \frac{i}{\sqrt{4\pi k}} \left( e^{-ik\pi R(t+z)} - e^{-ik\pi R(t-z)} \right). \quad (2)$$

The Dirichlet boundary condition is satisfied provided that $R(t)$ satisfies Moore's equation

$$R(t + L(t)) - R(t - L(t)) = 2. \quad (3)$$

Let us now extend this treatment to Neumann boundary conditions. In the instantaneous frame where the moving boundary is at rest they read $n^\mu \partial_\mu \phi|_{\text{boundary}} = 0$, where $n^\mu$ is a unit two-vector perpendicular to the trajectory of the boundary. The non-relativistic version of these boundary conditions in the laboratory frame read

$$\partial_z \phi|_{z=0} = 0 \; ; \; (\partial_z + \dot{L}\, \partial_t)\phi|_{z=L(t)} = 0. \quad (4)$$

The field inside the cavity can be expanded as a sum of modes

$$\phi(z,t) = A + B\phi_0(z,t) + \sum_{k=1}^{\infty} \left[ a_k \phi_k(z,t) + a_k^\dagger \phi_k^*(z,t) \right]. \quad (5)$$

The first two terms correspond to the quantization of the zero mode. The time-independent, hermitian operators $A$ and $B$ satisfy the commutation relations $[A,B] = i$, and $[A, a_k] = [B, a_k] = 0$. In analogy to the quantization of an open string [12], the operator $A$ corresponds to the initial position of the center of mass of the string, while the operator $B$ is associated to the average momenta of the string. The Hilbert space associated with the zero mode can be spanned by the eigenstates $|b\rangle$ of the momentum operator $B$: $B|b\rangle = b|b\rangle$. For $t < 0$ the zero mode function is position independent $\phi_0(t < 0) = t/L_0$. For $t > 0$, it can be written in terms of the Moore's function as

$$\phi_0(z,t) = \frac{1}{2}[R(t+z) + R(t-z)]. \quad (6)$$

The other mode functions $\phi_k(z,t)$ are positive frequency modes for $t < 0$, and satisfy instantaneous Neumann boundary conditions. Analogously to Eq.(2) for the Dirichlet case, these modes can be written in terms of Moore's function $R(t)$ as

$$\phi_k(z,t) = \frac{1}{\sqrt{4\pi k}} \left( e^{-ik\pi R(t+z)} + e^{-ik\pi R(t-z)} \right). \quad (7)$$

Note the change of sign between Eq.(2) for Dirichlet modes, and Eq.(7) for Neumann modes. Taking the time derivative of Moore's equation for $R(t)$ it is straightforward to prove that this expression for the modes automatically verifies Neumann boundary conditions Eq.(4) in the laboratory frame, both at $z = 0$ and $z = L(t)$.

The complete solution to the problem for both types of boundary conditions involves finding a solution $R(t)$ in terms of the prescribed motion $L(t)$. The modes are positive frequency modes for $t < 0$ if $R(t) = t/L_0$ for $-L_0 \leq t \leq L_0$, which is indeed a solution to Eq.(3) for $t < 0$. For $t > 0$, Eq.(3) can be solved, for example, using a renormalization group improvement of the perturbative solution [13]. Such an improvement is needed because the perturbative solution contains secular terms, and is therefore valid only for short times $\epsilon \Omega t \ll 1$. The RG-improved solution, valid for longer times $\epsilon^2 \Omega t \ll 1$, reads

$$R(t) = \frac{t}{L_0} - \frac{2}{\pi q} \text{Im} \ln \left[ 1 + \xi + (1-\xi) e^{\frac{iq\pi t}{L_0}} \right], \tag{8}$$

where $\xi = \exp[(-1)^{q+1} \pi q \epsilon t / L_0]$. The function $R(t)$ develops a staircase shape for long times [13, 14]. Within regions of $t$ between odd multiples of $L_0$ there appear $q$ jumps, located at values of $t$ satisfying $\cos(q\pi t/L_0) = \mp 1$, the upper sign corresponding to even values of $q$ and the lower one to odd values of $q$.

The energy density of the field is given by $\langle T_{00}(x,t) \rangle = -f(t+z) - f(t-z)$ [15], where

$$f = \frac{1}{24\pi} \left[ \frac{R'''}{R'} - \frac{3}{2} \left( \frac{R''}{R'} \right)^2 + \frac{\pi^2}{2} (R')^2 \right]. \tag{9}$$

This expression is valid both for Dirichlet and Neumann boundary conditions. The expectation value is taken in the vacuum state. In the Dirichlet case this vacuum state is annihilated by all $a_k$, i.e., $a_k|0\rangle = 0$. In the Neumann case the vacuum state is annihilated by both $B$ and $a_k$: $B|0\rangle = a_k|0\rangle = 0$. For other initial quantum states some differences may arise [16]. For example, when the initial state $|i\rangle$ satisfies $B|i\rangle = b|i\rangle$, and $a_k|0\rangle = 0$, in the Neumann case the mean value of the energy density contains an additional term proportional to $b^2[(R'(t+z))^2 + (R'(t-z))^2]$. In what follows we consider the $b = 0$ case. For $q = 1$ ("semi-resonant" case) no exponential amplification of the energy density is obtained, whereas for $q \geq 2$ ("resonant" cases) the energy density grows exponentially in the form of $q$ traveling wave packets which become narrower and higher as time increases. The number of created particles can be computed from the solution Eq.(8). Photons are created resonantly in all modes with $n = q + 2j$, with $j$ a non-negative integer. This is due to the fact that the spectrum of a one dimensional cavity is *equidistant*: although the external frequency resonates with a particular eigenmode of the cavity, intermode coupling produces resonant creation in the other modes. While the number of photons in each mode grows quadratically in time, the total energy inside the cavity grows exponentially. The spectrum of motion-induced photons is the same for both Dirichlet and Neumann boundary conditions. This can be confirmed by expressing the Bogoliubov coefficients in terms of the Moore's function, starting from Eqs.(2,7).

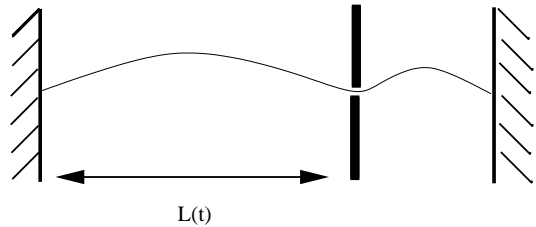

**Figure 1.** A mechanical analogue for the dynamical Casimir effect. A moving plate with a hole enforces the Dirichlet boundary condition at the position $L(t)$

## 3. A mechanical analogue of the 1+1 dynamical Casimir effect

It is well known that one can excite transversal waves in an elastic string by stretching it longitudinally. This is the famous Melde experiment, discussed in many elementary physics courses [17]. For a periodic stretching, there is a parametric resonance effect when the external frequency equals twice the frequency of an eigenmode of the string (the resonance makes non-linearities to become important as time goes on, but we will not discuss this issue here). That particular eigenmode is amplified resonantly. What is less known is that one can also excite transversal oscillations on the string by changing its longitude, even keeping a constant tension (see Fig. 1). This situation has been considered a long time ago to illustrate the radiation pressure: a string with both ends fixed that passes through a small hole on a plate that moves along the string [18]. We will use this classical mechanical system as a simulator of the dynamical Casimir effect.

Let us denote by $y(z,t)$ the transversal displacement of the string, and by $v_0$ the velocity of wave propagation. After rescaling time as $t \to v_0 t$, $y$ satisfies the wave equation $(\partial_{zz}^2 - \partial_{tt}^2)y = 0$, with Dirichlet boundary conditions at both ends of the string and at the position of the hole. As we assume the plate is a perfect reflector, we can consider independently the parts of the string which are the left or the right of the hole. At the classical level, the problem for each part of the string is equivalent to the 1+1 dynamical Casimir effect for a massless field. Therefore, we can solve for $y(z,t)$ using Moore's approach. Let us assume that for $t < 0$ the plate is fixed at the position $z = L_0$, and that the portion of the string $0 < z < L_0$ is vibrating in its eigenmode $k$, i.e., $y(z, t < 0) = 2a_k \sin(k\pi z/L_0) \cos(k\pi t/L_0)$. When the plate starts to move at $t > 0$, following a prescribed trajectory $L(t)$, the solution is

$$y(z, t > 0) = a_k \left[\sin(k\pi R(t+z)) - \sin(k\pi R(t-z))\right], \tag{10}$$

where $R(t)$ satisfies Moore's equation Eq.(3). Assuming that the plate oscillates around $L = L_0$ at a frequency $\Omega = 2k\pi/L_0$, the function $R(t)$ is explicitly given by Eq.(8) with $q = 2k$. Due to the staircase form of this function, one can readily check that the string develops $q$ peaks that bounce back and forth between the fixed end at $x = 0$ and the moving plate at $z = L(t)$ (see fig. 2). The energy density is of the form $\rho_E = T[g(t+z) + g(t-z)]$, where $T$ is the tension of the string and $g(t) = (a_k^2/2)\cos^2(k\pi R(t))\dot{R}^2(t)$. It also develops $q$ peaks as times goes on (see fig. 3).

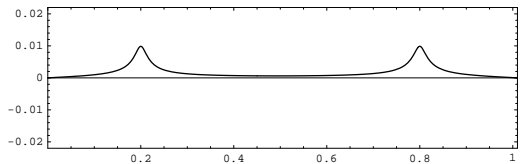

**Figure 2.** Pulses on the string induced by a time dependent length. We plot the transversal displacement $y(z,t)$ as a function of $z$ for a fixed time. We assume $k = 1$, $L = 1$m, $a_1 = 0.01$m, $v = 1$m/s, $q = 2$, and $t = 50.3$s.

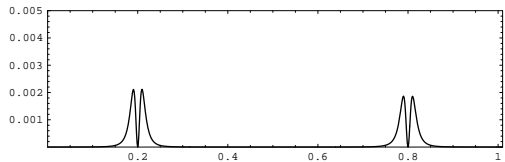

**Figure 3.** Pulses in the energy density of the string. We plot $\rho_E/T$ as a function of $z$ for a fixed time. The parameters are as in the previous figure. Each pulse has two peaks, the minimum between them corresponding to the maximum of $y(z,t)$ in fig.2. Note that at those points $\partial_t y = \partial_z y = 0$, and therefore $\rho_E = 0$

In order to excite transversal oscillations it is necessary to have an initial classical wave on the string. This "seed" is amplified by parametric resonance due to the motion of the plate. The dynamics of the string with time dependent length is completely different from the usual Melde experiment (in which the tension depends on time) since in that case only one mode is excited by parametric resonance. In the quantum case, starting from vacuum, there are no classical vibrations initially. However, the initial conditions of the modes are non trivial due to Heisenberg uncertainty principle. These initial conditions are amplified by parametric resonance, and the evolution of the modes produce a non trivial energy density inside the cavity. This simple analogy will allow us to understand why the dynamical Casimir effect is non trivial for spherical cavities and for cylindrical cavities with time dependent radius (see below).

## 4. TEM modes: a physical realization of the 1+1 dynamical Casimir effect

A cylindrical and non-simply connected cavity admit TEM modes, for which both the electric and magnetic fields have vanishing longitudinal components. Denoting by $z$ the coordinate along the axis of the cavity and by $\mathbf{x}_\perp$ the coordinates the perpendicular plane, the potential vector $\mathbf{A}$ for the TEM solutions is

$$\mathbf{A}(\mathbf{x}_\perp, z, t) = \mathbf{A}_\perp(\mathbf{x}_\perp)\varphi(z,t), \quad (11)$$

where $\mathbf{A}_\perp$ is a solution of an *electrostatic* problem in the two transverse dimensions. Note that this electrostatic solution does not exist in a simply connected cavity. The scalar field $\varphi$ satisfies Dirichlet boundary conditions on the longitudinal boundaries $z = 0$ and $z = L_z(t)$, and the longitudinal wave equation $(\partial_t^2 - \partial_z^2)\varphi = 0$. For a static cavity, the eigenfrequencies of the TEM modes are $w_n = n\pi/L_z$. Note that this is an *equidistant spectrum*. The Hamiltonian associated with TEM modes is

$$H^{\text{TEM}} = \frac{1}{8\pi} \int d^2x_\perp dz \, (\mathbf{E}^2 + \mathbf{B}^2) = \frac{1}{8\pi} \left( \int d^2x_\perp |\mathbf{A}_\perp|^2 \right) \int dz[(\partial_t\varphi)^2 + (\partial_z\varphi)^2]. \quad (12)$$

The above equations show that the quantization of TEM modes is equivalent to the quantization of a massless real scalar field in $1+1$ dimensions, satisfying Dirichlet boundary conditions at $z=0$ and $z=L_z(t)$. Therefore, the results of Section II describe the evolution of the electromagnetic energy density inside a non simply connected cavity with time dependent length. As we will see, this evolution is very different from those of the TE and TM modes.

## 5. 3D cavities: TE and TM modes

### 5.1. Hertz potentials formalism

When studying the quantization of the full electromagnetic field inside cavities with moving boundaries it is convenient to express the physical degrees of freedom of the electromagnetic field in terms of Hertz potentials [19, 20, 21]. The standard vector $\mathbf{A}$ and scalar $\Phi$ potentials are written in terms of the electric $\mathbf{\Pi}_e$ and magnetic $\mathbf{\Pi}_m$ vector Hertz potentials as $\Phi = -\frac{1}{\epsilon}\nabla \cdot \mathbf{\Pi}_e$ and $\mathbf{A} = \mu\frac{\partial \mathbf{\Pi}_e}{\partial t} + \nabla \times \mathbf{\Pi}_m$, where $\epsilon$ is the electric permittivity, and $\mu$ is the magnetic permeability. In vacuum, at points away from the sources, it is possible to write each of the vector Hertz potentials in terms of scalar Hertz potentials,

$$\mathbf{\Pi}_e = \phi^{\text{TM}} \, \hat{\mathbf{e}} \ ; \ \mathbf{\Pi}_m = \phi^{\text{TE}} \, \hat{\mathbf{e}} \, , \quad (13)$$

where $\hat{\mathbf{e}}$ is a unit vector. For example, in the case of waveguides, $\hat{\mathbf{e}}$ is the unit vector along the axis of the waveguide, say $\hat{\mathbf{z}}$, and for a spherical cavity it is the radial unit vector $\hat{\mathbf{r}}$. The field $\phi^{\text{TE}}$ gives rise to TE fields with respect to $\hat{\mathbf{e}}$, whereas $\phi^{\text{TM}}$ represents TM fields. Alternatively [22, 23], it is also possible to use two vector potentials $\mathbf{A}_{\text{TE}}$ and $\mathbf{A}_{\text{TM}}$ related to the above defined scalar Hertz potentials as $\mathbf{A}_{\text{TE}} = \hat{\mathbf{e}} \times \nabla\phi^{\text{TE}}$ and $\mathbf{A}_{\text{TM}} = \hat{\mathbf{e}} \times \nabla\phi^{\text{TM}}$. The EM field inside moving cylindrical or spherical cavities with perfect conducting moving boundaries can be described in terms of these two *independent* scalar Hertz potentials, as no mixed terms appear in Maxwell's Lagrangian or Hamiltonian. For cylindrical cavities, the scalar Hertz potentials satisfy the Klein Gordon equation. In the spherical case, the Klein Gordon equation is satisfied by the so called Debye potentials $\psi^{\text{TE,TM}} = \phi^{\text{TE,TM}}/r$.

The boundary conditions of the scalar Hertz potentials on the static, perfectly reflecting boundaries $S_{\text{static}}$ of the cavity are the usual Dirichlet boundary condition for $\phi^{\text{TE}}$, and the usual Neumann condition for $\phi^{\text{TM}}$. On the moving wall $S_{\text{mov}}$ these boundary conditions are first imposed on the instantaneous moving reference frame, and then Lorentz transformed to the laboratory frame. Denoting by $d(t)$ the coordinate of the moving wall, the boundary conditions read

$$\phi^{\text{TE}}\big|_{S_{\text{mov}}} = 0 \; ; \; (\partial_e + \dot{d}\,\partial_t)\phi^{\text{TM}}\big|_{S_{\text{mov}}} = 0. \tag{14}$$

### 5.2. Photon creation in cylindrical cavities

In this subsection we consider the dynamical Casimir effect inside closed cylindrical cavities, one of whose end caps is moving harmonically $L_z(t) = L_0[1 + \epsilon \sin(\Omega t)]$, with $\epsilon \ll 1$. We assume that the motion starts at $t = 0$ and stops at $t = T$. The field equations for both scalar Hertz potentials are $\Box \phi^{\text{TE}} = \Box \phi^{\text{TM}} = 0$, which should be solved under the time-dependent boundary conditions described above. The quantization procedure has been described in detail in previous papers [24, 9, 25]. Here we review the main results. At any given time $0 < t < T$ both scalar Hertz potentials can be expanded in terms of an instantaneous basis

$$\phi^{\text{TE,TM}}(\mathbf{x}, t) = \sum_{\mathbf{k}} a_{\mathbf{k}}^{\text{IN}} C_{\mathbf{k}}^{\text{TE,TM}} u_{\mathbf{k}}^{\text{TE,TM}}(\mathbf{x}, t) + \text{c.c.}, \tag{15}$$

where $a_{\mathbf{k}}^{\text{IN}}$ are bosonic operators that annihilate the IN vacuum state for $t < 0$, and $C_{\mathbf{k}}^{\text{TE,TM}}$ are normalization constants (these must be appropriately included to obtain the usual form of the electromagnetic Hamiltonian in terms of the electric and magnetic fields; see [25] for details). The mode functions are

$$u_{\mathbf{k}}^{\text{TE}} = \sum_{\mathbf{p}} Q_{\mathbf{p},\text{TE}}^{(\mathbf{k})}(t) \sqrt{2/L_z(t)} \sin\left(\frac{p_z \pi z}{L_z(t)}\right) v_{\mathbf{p}_\perp}(\mathbf{x}_\perp);$$

$$u_{\mathbf{k}}^{\text{TM}} = \sum_{\mathbf{p}} [Q_{\mathbf{p},\text{TM}}^{(\mathbf{k})}(t) + \dot{Q}_{\mathbf{p},\text{TM}}^{(\mathbf{k})}(t) g(z,t)] \sqrt{2/L_z(t)} \cos\left(\frac{p_z \pi z}{L_z(t)}\right) r_{\mathbf{p}_\perp}(\mathbf{x}_\perp).$$

Here the index $\mathbf{p} \neq 0$ is a vector of non-negative integers. Note that for the TM case the zero mode $\mathbf{p} = 0$ (that is $z$-independent and, in principle, is a possible solution to the field equations and boundary conditions along the $z$ direction) is not a solution for the 3D problem we are considering, since it does not satisfy the boundary conditions on the static surfaces. However, this zero mode (associated with Neumann boundary conditions) is present in the case of the 1+1 dynamical Casimir effect, as was discussed in Section 2. The function $g(z,t) = \dot{L}_z(t) L_z(t) \xi(z/L_z(t))$ (where $\xi(z)$ is a solution to the conditions $\xi(0) = \xi(1) = \partial_z \xi(0) = 0$, and $\partial_z \xi(1) = -1$) appears when expanding the TM modes in an instantaneous basis and taking the small $\epsilon$ limit. There are many solutions for $\xi(z)$, but all of them can be shown to lead to the same physical results [9]. The mode functions $v_{\mathbf{p}_\perp}(\mathbf{x}_\perp)$ and $r_{\mathbf{p}_\perp}(\mathbf{x}_\perp)$, are described below for different types of cavities.

The mode functions $Q_{\mathbf{p},\text{TE/TM}}^{(\mathbf{k})}$ satisfy second order, mode-coupled linear differential equations

$$\ddot{Q}_{\mathbf{p},\text{TE}}^{(\mathbf{k})} + \omega_{\mathbf{p}}^2(t) Q_{\mathbf{p},\text{TE}}^{(\mathbf{k})} = 2\lambda(t) \sum_{\mathbf{j}} g_{\mathbf{p}\mathbf{j}} \dot{Q}_{\mathbf{j},\text{TE}}^{(\mathbf{k})} + \dot{\lambda}(t) \sum_{\mathbf{j}} g_{\mathbf{p}\mathbf{j}} Q_{\mathbf{j},\text{TE}}^{(\mathbf{k})} + O(\epsilon^2), \tag{16}$$

and

$$\ddot{Q}^{(\mathbf{k})}_{\mathbf{p},\text{TM}} + \omega^2_{\mathbf{k}}(t) Q^{(\mathbf{k})}_{\mathbf{p},\text{TM}} = -2\lambda(t) \sum_{\mathbf{j}} h_{\mathbf{jp}} \dot{Q}^{(\mathbf{k})}_{\mathbf{p},\text{TM}} - \dot{\lambda}(t) \sum_{\mathbf{j}} h_{\mathbf{jp}} Q^{(\mathbf{k})}_{\mathbf{p},\text{TM}}$$
$$- 2\dot{\lambda}(t) L_z^2(t) \sum_{\mathbf{j}} s_{\mathbf{jp}} \ddot{Q}^{(\mathbf{k})}_{\mathbf{p},\text{TM}} - \sum_{\mathbf{j}} \dot{Q}^{(\mathbf{k})}_{\mathbf{p},\text{TM}} [s_{\mathbf{jp}} \ddot{\lambda}(t) L_z^2(t) - \lambda(t) \eta_{\mathbf{jp}}]$$
$$- \lambda(t) L_z^2(t) \sum_{\mathbf{j}} s_{\mathbf{jp}} \partial_t^3 Q^{(\mathbf{k})}_{\mathbf{p},\text{TM}} + O(\epsilon^2). \quad (17)$$

In these equations, $\lambda(t) = \dot{L}_z(t)/L_z(t)$, $\omega_{\mathbf{p}}(t) = \sqrt{\mathbf{p}_\perp^2 + (p_z\pi/L_z(t))^2}$, and the coupling coefficients $g_{\mathbf{jp}}$, $s_{\mathbf{jp}}$, $\eta_{\mathbf{jp}}$, and $h_{\mathbf{jp}}$ are defined in [9]. The above equations can be solved using different approximation methods, like multiple scale analysis, which is described in our previous works [24, 9]. For the "parametric resonant case" ($\Omega = 2\omega_{\mathbf{k}}$ for some $\mathbf{k}$) the solutions present resonant behavior (*i.e.*, exponential growth). Moreover, for some particular geometries and sizes of the cavities, different modes $\mathbf{j}$ and $\mathbf{k}$ can be coupled, provided either of the resonant coupling conditions $\Omega = |\omega_{\mathbf{k}} \pm \omega_{\mathbf{j}}|$ are met. When intermode coupling occurs it affects the rate of photon creation, typically resulting in a reduction of that rate.

The solutions to the mode equations provide us with expressions of the Bogoliubov coefficients $A^{\mathbf{k}}_{\mathbf{p},\text{TE/TM}}$ and $B^{\mathbf{k}}_{\mathbf{p},\text{TE/TM}}$ that relate the IN basis ($t < 0$, before the motion starts) and the OUT basis ($t > T$, after the motion ends). The number of motion-induced photons with a given wavevector $\mathbf{k}$ and polarization TE or TM can be calculated in terms of the Bogoliubov coefficients $B^{\mathbf{k}}_{\mathbf{p},\text{TE/TM}}$. Except for special geometries, in general the resonant coupling conditions are not met: different $\mathbf{k}$ modes will not be coupled during the dynamics, and Eq.(16) and Eq.(17) reduce to the Mathieu equation for a single mode. In consequence, the number of motion-induced photons in that given mode will grow exponentially. The growth rate is different for TE and TM modes

$$\langle N_{\mathbf{k},\text{TE}}(t)\rangle = \sinh^2(\lambda_{\mathbf{k},\text{TE}}\epsilon t) \; ; \; \langle N_{\mathbf{k},\text{TM}}(t)\rangle = \sinh^2(\lambda_{\mathbf{k},\text{TM}}\epsilon t), \quad (18)$$

where $\lambda_{\mathbf{k},\text{TE}} = k_z^2/2\omega_{\mathbf{k}}$ and $\lambda_{\mathbf{k},\text{TM}} = (2\omega_{\mathbf{k}}^2 - k_z^2)/2\omega_{\mathbf{k}}$. When both polarizations are present, the rate of growth for TM photons is larger than for TE photons, *i.e.*, $\lambda_{\mathbf{k},\text{TM}} > \lambda_{\mathbf{k},\text{TE}}$.

*Rectangular section:* For a waveguide of length $L_z(t)$ and transversal rectangular shape (lengths $L_x, L_y$), the TE mode function is

$$v_{n_x,n_y}(\mathbf{x}_\perp) = \frac{2}{\sqrt{L_x L_y}} \cos\left(\frac{n_x \pi x}{L_x}\right) \cos\left(\frac{n_y \pi y}{L_y}\right), \quad (19)$$

with $n_x$ and $n_y$ non-negative integers that cannot be simultaneously zero. The TM mode function is

$$r_{m_x,m_y}(\mathbf{x}_\perp) = \frac{2}{\sqrt{L_x L_y}} \sin\left(\frac{m_x \pi x}{L_x}\right) \sin\left(\frac{m_y \pi y}{L_y}\right), \quad (20)$$

where $m_x, m_y$ are integers such that $m_x, m_y \geq 1$. The spectrum is

$$\omega_{n_x,n_y,n_z} = \sqrt{(n_x\pi/L_x)^2 + (n_y\pi/L_y)^2 + (n_z\pi/L_z)^2}, \quad (21)$$

As an example, let us analyze the case of a cubic cavity of size $L$ under the parametric resonant condition $\Omega = 2\omega_\mathbf{k}$. The fundamental TE mode is doubly degenerate ($(1, 0, 1)$ and $(0, 1, 1)$) and uncoupled to other modes. The fundamental TM mode has the same energy as the fundamental TE mode, and it is coupled to the TM mode $(1, 1, 4)$. Motion-induced TM photons are produced exponentially as $\exp(\pi\epsilon t/\sqrt{2}L)$, and much faster than TE photons.

*Circular section:* For a waveguide of length $L_z(t)$ and transversal circular shape (radius $R$), the TE mode function is

$$v_{nm}(\mathbf{x}_\perp) = \frac{1}{\sqrt{\pi}} \frac{1}{RJ_n(y_{nm})\sqrt{1 - n^2/y_{nm}^2}} J_n\left(y_{nm}\frac{\rho}{R}\right) e^{in\phi}, \qquad (22)$$

where $J_n$ denotes the Bessel function of $n$th order, and $y_{nm}$ is the $m$th positive root of the equation $J_n'(y) = 0$. The eigenfrequencies are given by

$$\omega_{n,m,n_z} = \sqrt{\left(\frac{y_{nm}}{R}\right)^2 + \left(\frac{n_z\pi}{L_z}\right)^2}, \qquad (23)$$

where $n_z \geq 1$. The TM mode function is

$$r_{nm}(\mathbf{x}_\perp) = \frac{1}{\sqrt{\pi}} \frac{1}{RJ_{n+1}(x_{nm})} J_n\left(x_{nm}\frac{\rho}{R}\right) e^{in\phi}, \qquad (24)$$

where $x_{nm}$ is the $m$th root of the equation $J_n(x) = 0$. The spectrum is given by Eq.(23) with $y_{nm}$ replaced by $x_{nm}$ and $n_z \geq 0$. Denoting the modes by $(n, m, n_z)$, the lowest TE mode is $(1, 1, 1)$ and has a frequency $\omega_{111} = (1.841/R)\sqrt{1 + 2.912(R/L_z)^2}$. This mode is uncoupled to any other modes, and the number of photons in this mode grows exponentially in time as $\exp\left(\pi\epsilon t/\sqrt{1 + 0.343(L_z/R)^2}L_z\right)$ when parametrically excited. The lowest TM mode $(0, 1, 0)$ is also uncoupled and has a frequency $\omega_{010} = 2.405/R$. The parametric growth is $\exp\left(4.81\epsilon t/R\right)$. For $L_z$ large enough ($L_z > 2.03R$), the resonance frequency $\omega_{111}$ of the lowest TE mode is smaller than that for the lowest TM mode. Then the $(1, 1, 1)$ TE mode is the fundamental oscillation of the cavity.

### 5.3. Photon creation in spherical cavities

For the case of fields inside a spherical cavity with time dependent radius $a(t) = a_0[1 + \epsilon \sin(\Omega t)]$, the Debye potentials $\psi^{\text{TE,TM}}$ satisfy the Klein Gordon equation. The mode functions can be written as

$$u_{k\ell m}^{\text{TE}} = \sum_p Q_{p,\text{TE}}^{(k)}(t)\sqrt{\frac{2}{a^3(t)}} \frac{1}{j_\ell'(j_{\ell p})} j_\ell(\frac{j_{\ell p}}{a(t)}r) \Psi_{\ell m}(\theta, \phi), \qquad (25)$$

and

$$u_{k\ell m}^{\text{TM}} = \sum_p [Q_{p,\text{TM}}^{(k)}(t) + \dot{Q}_{p,\text{TM}}^{(k)}(t)g(r,t)]\phi_{p\ell m}(\mathbf{r}, a(t)), \qquad (26)$$

with

$$\phi_{p\ell m}(\mathbf{r}, a(t)) = \sqrt{\frac{2}{a^3(t)}} \frac{1}{\jmath'_\ell(\kappa_{\ell p})} \frac{1}{\sqrt{\kappa_{\ell p}^2 - \ell(\ell+1)}} \jmath_\ell(\frac{\kappa_{\ell p}}{a(t)} r) \Psi_{\ell m}(\theta, \phi). \quad (27)$$

The function $g(r,t)$ can be expressed as $g(r,t) = \dot{a}(t)a(t)v(r/a(t))$, where $v(1) = 0$ and $v'(1) = -1$. In Eqs. (25) and (27), $\Psi_{\ell m}(\theta, \phi)$ denote the usual spherical harmonics, $\jmath_{\ell k}$ is the k-th zero of the spherical function $\jmath_\ell(x)$, and $\kappa_{\ell k}$ is the k-th zero of $\partial_x[x\jmath_\ell(x)] = 0$. The frequencies of the modes are given by

$$\omega_{\ell k}^{\text{TE}} = \frac{\jmath_{\ell k}}{a_0} \quad ; \quad \omega_{\ell k}^{\text{TM}} = \frac{\kappa_{\ell k}}{a_0}. \quad (28)$$

The equations of motion for the time dependent coefficients $Q_{p,\text{TE}}^{(k)}$ and $Q_{p,\text{TM}}^{(k)}$ can be obtained as in the previous cases [26]. For $\ell = 0$ the spectrum is equidistant, and the associated spherically symmetric modes, present for the scalar fields $\psi^{\text{TE,TM}}$, do not contribute to the full electromagnetic field because the vector potentials $\mathbf{A}_{\text{TE}}$ and $\mathbf{A}_{\text{TM}}$ are obtained by applying the operator $\mathbf{r} \times \nabla$. This is a manifestation of the fact that there is no classical electromagnetic radiation with spherical symmetry. For $\ell \neq 0$, the spectrum for both TE and TM modes is not equidistant, and there is no mode coupling for the lowest frequencies. For the resonant case $\Omega = 2\omega_{\ell n}$ ($\ell \neq 0$) the number of photons created grows as $\exp(2\gamma\epsilon t)$, where

$$\gamma_{\text{TE}} = \frac{\omega_{\ell n}^{\text{TE}}}{2} \quad ; \quad \gamma_{\text{TM}} = \frac{\omega_{\ell n}^{\text{TM}}}{2} \frac{1 + \frac{\ell(\ell+1)}{\kappa_{\ell n}^2}}{1 - \frac{\ell(\ell+1)}{\kappa_{\ell n}^2}}. \quad (29)$$

Although there are no degenerate TE and TM modes, it is interesting to remark that $\gamma_{\text{TE}}/\omega_{\ell n}^{\text{TE}} < \gamma_{\text{TM}}/\omega_{\ell n}^{\text{TM}}$, as in the case of cylindrical cavities.

The mean value of the total angular momentum of the electromagnetic field inside the cavity vanishes in the vacuum state. Due to the spherical symmetry of the problem it must be zero for all times. It is possible to check that photons are created in singlet states [26]. Because of Gauss law, the classical electromagnetic field does not contain a time dependent $\ell = 0$ mode. Therefore, the assumption of spherical symmetry kills any classical radiation. However, at the quantum level, all modes with $\ell \neq 0$ have non vanishing fluctuations, which act as "seeds" for the dynamical Casimir effect. A similar argument applies to the case of non vanishing photon creation in cylindrical cavities with time dependent radius [27].

## 6. Conclusions

Motion induced radiation has a broad interest in the context of quantum field theory under the influence of external conditions. The ongoing challenging projects to demonstrate experimentally this tiny effect deserve theoretical efforts to compute it in different situations and with different approximations. In this paper we calculated the resonant photon creation for cylindrical and spherical cavities with time dependent sizes. We obtained the photon creation rates for TE and TM modes. We also considered the dynamical Casimir effect in $1+1$ dimensions, both with Dirichlet and Neumann boundary conditions, and showed that this toy model can be realized physically by TEM modes in a coaxial cylindrical waveguide.

## Acknowledgments

We are grateful to Prof. E. Elizalde for the kind invitation to attend QFEXT'05. We thank M. Crocce, F.C. Lombardo, and M. Ruser for fruitful discussions. The work of FDM and XOM has been supported by CONICET, UBA and ANPCyT, Argentina.